# Framework for Adoption of Generative Artificial Intelligence (GenAI) in Education

Samar Shailendra, Rajan Kadel, and Aakanksha Sharma

*SITE, Melbourne Institute of Technology, Australia*

*Abstract*—**Contributions: An adoption framework to include GenAI in the university curriculum. It identifies and highlights the role of different stakeholders (university management, students, staff, etc.) during the adoption process. It also proposes an objective approach based upon an evaluation matrix to assess the success and outcome of the GenAI adoption.**

**Background: Universities worldwide are debating and struggling with the adoption of GenAI in their curriculum. GenAI has impacted our perspective on traditional methods of academic integrity and the scholarship of teaching, learning, and research. Both the faculty and students are unsure about the approach in the absence of clear guidelines through the administration and regulators. This requires an established framework to define a process and articulate the roles and responsibilities of each stakeholder involved.**

**Research Questions: Whether the academic ecosystem requires a methodology to adopt GenAI into its curriculum? A systematic approach for the academic staff to ensure the students' learning outcomes are met with the adoption of GenAI. How to measure and communicate the adoption of GenAI in the university setup?**

**Methodology: The methodology employed in this study focuses on examining the university education system and assessing the opportunities and challenges related to incorporating GenAI in teaching and learning. Additionally, it identifies a gap and the absence of a comprehensive framework that obstructs the effective integration of GenAI within the academic environment.**

**Findings: The literature survey results indicate the limited or no adoption of GenAI by the university, which further reflects the dilemma in the minds of different stakeholders. For the successful adoption of GenAI, a standard framework is proposed i) for effective redesign of the course curriculum, ii) for enabling staff and students, iii) to define an evaluation matrix to measure the effectiveness and success of the adoption process.**

## I. INTRODUCTION

Artificial Intelligence (AI) has permeated our lives unlike any other technology in recent history. Generative AI (GenAI) has revolutionised how we approach education and deliver it to the students. There is a sense of uncertainty among the academicians regarding the role and integration of GenAI in providing education to university students. While many educators are hesitant to embrace and incorporate GenAI into their teaching materials and pedagogy, this paper contends that adopting it offers significant advantages in enhancing the quality of education. However, a structured approach is essential for the effective adoption of GenAI in the education system.

The key contributions of this paper can be summarised as:
- A holistic framework is introduced for adopting GenAI in education, considering the governance process, curriculum development, and stakeholders.
- The proposed framework delineates the outcome(s) of each stage, the flow of information, and the documentation involved between these stages.
- The proposed framework tackles various issues raised by GenAI, including academic integrity, privacy, ethics, and the uncertainties associated.

Through a thorough exploration of opportunities and challenges and an in-depth review of the literature, this paper aims to elucidate the transformative potential of GenAI in education.

This paper delineates three primary concerns stemming from introducing GenAI into education. These concerns encompass the need for i) A clear vision and direction regarding GenAI's utilisation, ii) The challenges related to implementing GenAI in teaching, learning, and research, and iii) The uncertainties and ethical considerations surrounding its application.

According to the United Nations Educational, Scientific, and Cultural Organisation (UNESCO), the use of GenAI in education must be regulated to harness its potential benefits [1]. Therefore, it is imperative for university management to provide a well-defined direction and vision within their policy framework to integrate GenAI into educational settings effectively. However, university management often grapples with delivering this clear direction due to the inherent uncertainties and associated risks. Hence, the foremost concern is the absence of a clear direction and vision for using GenAI in educational settings.

The implementation challenges tied to GenAI in teaching, learning, and research give rise to multiple obstacles. The primary challenge centres on the preservation and improve students' learning objectives, cognitive skills, and critical thinking abilities. The second challenge involves educators comprehending the technology and its implications for curriculum development, encompassing assessment design. The third challenge is the potential over-reliance on technology, which may reduce student motivation and hinder the achievement of learning outcomes [2]. Another significant challenge revolves around addressing academic integrity, privacy, and ethical concerns in the broader educational and research contexts when GenAI is employed. For example, there is uncertainty about how students and staff will respond to the integration of GenAI in education and the potential repercussions of inaccurate responses from GenAI products. We acknowledge that many concerns related to the adoption of GenAI are valid, in this paper, we not only highlight these concerns but also propose a framework to analyse and effectively address them.

The outline of the paper is as follows: Section II provides the opportunities and challenges associated with GenAI in education. Section III provides a brief review of frameworks in the existing literature on the adoption of GenAI in education and research gaps. Section IV outlines our framework development approach. The proposed framework for the adoption of GenAI in education is introduced in Section V. Finally, Section VI presents the conclusion and future directions.

## II. GenAI - Opportunities and Challenges

In recent years, educational institutions have recognised the importance of integrating GenAI into their educational practices, offering AI courses based on a theoretical foundation that includes machine learning, natural language processing, computer vision, and image processing. GenAI can perform a range of tasks, such as essay writing, summarising literature reviews, translation, paraphrasing, and generating assessments, making it increasingly popular in education [3].

### A. Opportunities with GenAI

GenAI offers numerous opportunities for the education sector by revolutionising a wide range of technology tools and software used in online learning, academic research, teaching assistance, exam evaluations, feedback on assignments, content creation etc., leading to enhanced students' educational experience [3]–[5].

*1) Providing Teaching Assistance:* GenAI can generate informative content and create presentations by suggesting templates, images, and graphics, including recommendations for enhancing engagement through these presentations, considering the target audience [6]. It provides teachers the ability to generate quizzes or tests and assist them in evaluating and providing valuable feedback. Additionally, it can address common questions and assist teachers in understanding complex problems [7]. Furthermore, with GenAI, integrating simulations and virtual laboratories enhances practical learning experiences and accessibility features, including text-to-speech and its reverse counterpart, making educational content more inclusive. By leveraging data analytics, educators can pinpoint student performance trends and patterns [7], facilitating the adjustment of teaching methods.

*2) Providing Research Assistance:* Researchers can briefly describe their research ideas to GenAI to seek benefits. GenAI can suggest research methods and provide various examples of how these methods were used in past studies. This assistance helps researchers better understand the methods and enhances their research inclusion [6]. GenAI can also assist in summarising the literature review, a task that may otherwise consume a significant amount of time, and it can offer suggestions for further study extensions. This guidance is very helpful and provides a clear direction for advancing research [3].

*3) Providing Writing Assistance:* GenAI aids in improving writing skills by generating content and providing feedback on sentence structure, grammatical issues, and punctuation. It offers suggestions for organising ideas and improving the flow of content. GenAI can also create compelling arguments by providing various examples of writing styles [6]. It can adjust sentence length, allowing users to request content on a topic in either 1000 words or a 200-words brief. GenAI's quick and valuable assistance is highly beneficial in education.

*4) Personalised Learning and Evaluation:* GenAI helps students in their learning by providing tutoring or homework help. Teachers can design custom exams for each student based on their specific learning requirements, thus evaluating student learning and progress [8]. Incorporating GenAI in education enhances student engagement, as they receive content that aligns with their unique learning styles and educational needs [6]. AI algorithms may analyse the student learning styles and adapt the curriculum to suit each student's needs, ensuring a customised educational experience. The personalised approach to learning and evaluation contributes significantly to academic success [9], [10].

### B. Challenges with GenAI

While GenAI has the potential to offer numerous advantages, it also presents many issues for its inclusion in the academic ecosystem. Its increased usage and alignment with students' learning outcomes pose significant challenges.

*1) Academic Integrity:* The foremost challenge is the responsible use of GenAI while considering ethical issues and maintaining academic integrity. Exposing GenAI to students increases the risk of plagiarism and affects academic integrity [11]. Students may misuse it to write essays or create solutions and claim them as their own work. This contradicts the principles of honesty and originality in academia [2]. Additionally, students may miss the opportunity to engage with the content and develop critical thinking skills. Addressing these challenges requires a balance between harnessing the benefits of GenAI in assessment design and implementing robust measures to prevent and detect plagiarism. It underscores the importance of promoting the ethical use of GenAI in education while preserving integrity and authenticity [8], [12], [13].

*2) Privacy:* With GenAI, privacy concerns are significantly higher because GenAI learns using a vast amount of raw data, which may include personal data as well [14]. Safe and proper usage of personal data is essential to protect privacy [15]. These systems are capable of creating fake identities to be used for malicious activities such as scams, identity theft, creation of fake audio and videos etc. In the education sector, privacy concerns are particularly alarming due to the involvement of students' personal data. To safeguard the student's privacy, it is crucial to ensure the secure and responsible usage of data.

In April of 2023, Italy made history by becoming the first nation to prohibit the GenAI model (ChatGPT) due to concerns related to privacy. The data protection authority confirmed that there was no legal foundation for collecting and using personal data in the training of ChatGPT [6], [13].

*3) Lack of Regulation:* As AI technologies become more prevalent in the learning environment, various issues emerge, including the need for clear guidelines [10], [16], [17]. Currently, there is no regulatory framework for GenAI, and each educational institution responds differently, leading to variations in the adoption of such technologies. This lack of standardisation can result in disparities in the quality of educational experiences and potential privacy risks for students [2]. The absence of established standards for content creation and assessment design using GenAI poses challenges in ensuring content alignment with educational needs, and therefore, clear regulations can address these concerns [13].

*4) Assessment Design and Evaluation Challenges:* With the incorporation of GenAI, assessment design and its evaluation are notable challenges. If these models are not well-trained and monitored, there is a risk of unfairness, non-transparency, and biases. Another significant challenge is to ensure that the submitted assessments truly represent the students' work. If students excessively rely on GenAI to complete their assessments, achieving the intended learning outcomes becomes challenging. To address these challenges, clear rules and guidelines are essential to confirm the fairness of the assessments and to ensure that students genuinely understand their assessment tasks [6], [10], [13].

*5) Cognitive Bias:* Ethical principles do not inherently govern GenAI; these models cannot differentiate between right and wrong, potentially leading to unfair outcomes. If the data contains biases or unfairness, the content generated by GenAI can also exhibit bias and unfairness. Moreover, these models may produce similar content for all students and may not adapt well to changing data patterns. To address these issues and reduce the risk of perpetuating biases and unfairness, GenAI models must be trained on fair and unbiased data [2], [13]. However, ensuring that data is fair and free from any biases is a significantly challenging task in education.

*6) Diversity and Inclusion:* The inclusion of GenAI poses significant challenges pertaining to diversity and inclusion in the education ecosystem. It is essential that GenAI understands and caters to the needs of students with disabilities or those who may not comprehend the English language, which causes them to be left behind. Hence, GenAI should be accessible in all languages and adaptable to meet the diverse learning needs of students [13].

*7) Accessibility:* While GenAI offers enhanced learning opportunities, these tools may not be universally accessible. GenAI may remain inaccessible in several countries due to government regulations etc. Even in countries that permit the use of GenAI, equal internet access for everyone remains a challenge. As a result, equal accessibility for everyone is questionable [2]. In the education sector, some staff familiar with GenAI can provide better guidance to their students compared to others unfamiliar with it. This discrepancy can create educational inequalities [13]. Hence, it is imperative to provide proper training for staff and students to use GenAI in education effectively.

*8) Hallucination:* The integration of GenAI into education introduces a significant challenge known as data hallucination, wherein AI confidently generates content that is non-factual and incorrect. This can occur due to various factors such as noisy data, erroneous parametric knowledge, flawed attention mechanisms, or inadequate training strategies [18]. Data hallucination in education is a concerning challenge as it can negatively impact students' understanding and learning outcomes. Within our own experience as well, students have submitted wrong or conflicting information generated by GenAI for their assignments. Implementing effective mitigation strategies, such as custom GenAI tools using quality & authentic data, rigorous data pre-processing, etc., is crucial to ensure the reliability and trustworthiness of AI-generated educational content [19]. Subsequently, expert review and cross-referencing can further reduce the inaccuracies introduced by GenAI.

The following section presents a survey of related literature on existing frameworks for the adoption of GenAI in education, highlighting their strengths/weaknesses and identifying any gaps to facilitate the successful adoption of GenAI in education.

## III. RELATED WORKS

A few works in the literature exist on the framework for using GenAI in education. In [17], the AI ecological education policy framework is introduced. It has organised the policy into three dimensions: governance, operational, and pedagogical. The governance dimension covers four key areas (academic integrity, ethical issues, AI governance, attributing AI technologies, and equity in accessing AI technologies) and is managed by the institute's senior management. The operational dimension covers two areas (monitoring and evaluation of AI implementation, as well as training and support for all stakeholders) and is managed by the teaching and learning team and the institute's IT team. The pedagogical dimension encompasses four key areas and is managed by teachers. These areas involve assessment design, preparing students for AI-driven workplaces, using a balanced approach to AI adoption, and developing students' holistic competency skills. However, there is a lack of clear interaction between the three dimensions in the framework, which poses challenges for practical implementation.

In [20], a framework called the "AI literacy model" is introduced for curriculum development in AI courses for undergraduate degrees. The model consists of five stages: Enable AI, Understand AI, Apply AI, Evaluate and Create AI, and AI Ethics. Each stage is accompanied by a description, the level of AI integration within the stage, and the associated learning outcomes. The proposed model is closely linked with four initiatives at the institute: curriculum development; academic program and pathways; AI undergraduate scholars and AI undergraduate medallion programs; and AI career development and industry engagement. This framework primarily focuses on the development of new AI-based curricula but lacks a connection to policy, academic integrity, and several other challenges associated with the use of AI in education. Furthermore, the interplay and communication between the five stages of the framework are not clearly defined.

In [10], the paper introduces a framework called the "IDEE framework" for utilising ChatGPT in education. The frame-work comprises four stages: Identify the desired outcomes; determine the appropriate level of automation; ensure ethical considerations; and evaluate the effectiveness. The paper presents one application example of the proposed framework. However, there is a lack of clear interaction among the four stages in the framework, which presents challenges for practical implementation.

In [21], the authors introduce the integration of GenAI into the Technological Pedagogical Content Knowledge (TPACK) framework, discussing its impact and roles in technological knowledge, pedagogical knowledge, content knowledge, and contextual knowledge. The study primarily focuses on the effects of the introduction of GenAI on the overall educational setting. However, the framework does not delve into the practical implementation aspects of the framework in a real-world setting. Therefore, a brief review of the literature indicates a need for a holistic framework that encompasses the entire process of the university education system, curriculum development, and teaching.

## IV. Framework Primer

The approach used to create the proposed conceptual framework is illustrated in Fig. 1. It took into account university administration procedures, challenges, and opportunities associated with GenAI, along with other critical factors for the development of the framework. The first step was to grasp the issues introduced by GenAI in the educational environment and identify the research problems. This step succeeded by concurrent efforts in analysing university policies & procedures, alongside conducting a thorough literature review on existing frameworks for GenAI adoption in the educational ecosystem. The policies and procedures for university administration vary from one university to another. There is a clear distinction between institutional governance, which oversees overall institutional management (policy, vision, mission and finance), and academic governance, which focuses on academic affairs (academic decision, academic quality assurance, student learning assurance, academic risk) [22]. Additionally, the findings by the Australasian Academic Integrity Network (AAIN) through institutional responses on the use of GenAI in the Australian universities [23] and UNESCO guidelines on AI and education for policymakers [16] were also considered during the policy analysis.

The literature review mainly focused on three areas: challenges and opportunities associated with GenAI in educational settings; existing work on frameworks for the adoption of GenAI in education; and factors to be considered when developing a conceptual framework. A comprehensive literature review on challenges and opportunities associated with GenAI in educational settings is conducted in Section II. A review of existing frameworks for the adoption of GenAI is presented in Section III, and the findings suggest a need for a comprehensive framework covering the entire university education system, curriculum development, and teaching & learning.

It's challenging to follow a universal set of processes for framework development that applies across all contexts. Each institute's educational landscape is a complex blend of traditions, curriculum development frameworks, policy priorities, human capabilities, and financial resources. Likewise, every institute's curriculum has unique strengths and weaknesses, rooted in its educational philosophy and approach. Most institutes have established either implicit or explicit educational priorities to guide the curriculum's direction. The relevant university policy and regulatory requirements need to be referred for detailed information. Therefore, adhering to a standard process for framework development is quite involved due to the absence of universal agreement on the processes and procedures [1], [16], [17], [20], [24].

Thus, the development of this conceptual framework is primarily guided by the key factors crucial for its successful adoption and implementation. According to the International Bureau of Education, UNESCO [24], the first factor is that the conceptual framework should be flexible to adapt as per the institute and its priorities. The second factor is that the framework should offer a systematic approach to organising and managing content (policies and procedures) to facilitate content generation [25]. Therefore, it should define the parameters, directions, and standards for curriculum policy and practice, ensuring that content development aligns with the specific requirements of the institution. The framework is central to the education system and can influence various curriculum policy and practice aspects. Therefore, the third factor is that the framework should engage with all stakeholders for consultation and should broadly be supported by the top management for a successful adoption [26]. The final factor of the conceptual framework is that it should be structured as a step-wise process with a logical sequence, and it should support the monitoring and evaluation of the framework to facilitate further improvements [10], [17], [27], [28].

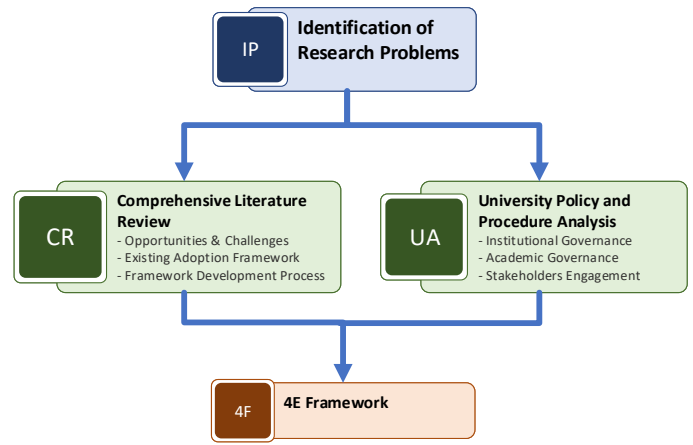

Fig. 1. Framework development approach.

The framework for adoption of GenAI in the education must address the following points [1], [9], [17], [28], [29]:

- Implementation strategies where university administrators, curriculum developer, and academics should align their perspectives to achieve the full potential of GenAI,
- Integration of GenAI with existing educational structures and practices,

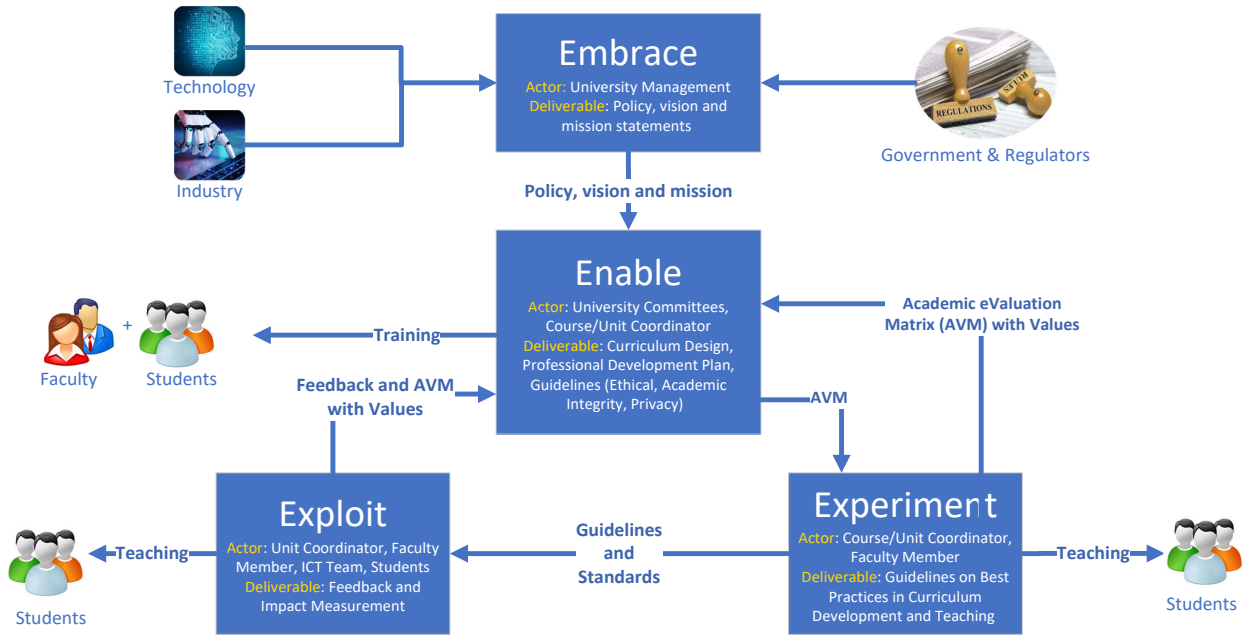

Fig. 2. **4E** - An adoption framework of GenAI in Education.

- Faculty/Students professional development in GenAI,
- Systems for student engagement and support for GenAI,
- Mechanism to ensure the proper flow of information among different bodies in the university to ensure proper communications and avoid misunderstanding,
- Uncertainty associated with impact on the use of GenAI on students and staff,
- Input from stakeholders (regulators, government agencies, and professional standard bodies, etc.) regarding GenAI, and
- Concerns raised by GenAI including academic integrity, privacy, ethics, inclusion, accessibility and cognitive bias.

Next, we will introduce a comprehensive framework for the adoption of GenAI in education by addressing the aforementioned aspects.

## V. **4E** - A GenAI Adoption Framework

GenAI has made significant progress and created a profound impact on our lives in recent times. However, experts contend that we are only scratching the surface of its capabilities. This paper delves into the imperative for education to not merely view GenAI as a technology but as an essential tool for nurturing our future generations. This mindset will enable the education system to empower the next generation to harness the boundless potential of GenAI effectively.

Despite the rapid advancements, the utilisation of GenAI is still in its early stages. It is crucial for everyone in the academic environment to comprehend, enable themselves, and adopt the role of GenAI appropriately. We argue that the adoption of GenAI in the academic curriculum is imperative [30]. A well-planned adoption will help maintain the ethics and scholarship of teaching and learning [31]. In contrast, an unplanned penetration is likely to cause panic, unfairness, and disparity among students and educators. Educators can have the potential to expose students to GenAI tools and utilise GenAI to evaluate students' performance. This approach is likely to provide the following outcomes:

1) Enables students to use GenAI, making them more innovative and competitive at the same time.
2) Prepares a level ground for everyone. No student is less or more equipped with GenAI resources. This will further provide an opportunity for educators to evaluate and compare performance, which is an essential aspect of academic learning.
3) Expected to bring out the best of the students and the faculty. They are more likely to innovate and contribute to the scholarship of teaching and learning.

In this paper, we identify and categorise the potential stakeholders involved during the process of adopting GenAI within the education framework.

### A. Stakeholders

It is evident that no other technology has revolutionised the academic system like GenAI. The regulators and authorities are brainstorming significantly over the role of GenAI in education [1], [23]. In the education ecosystem, there are five major stakeholders who need to align, enable and work in unison for the successful adoption of GenAI [24], [32], [33].

*1) **Management**:* This includes the decision makers such as University management, representatives from the Academic Board and Learning & Teaching committee, etc. They have similar goals: imparting quality education to students, increasing student enrolment, growing revenue, preventing malpractices, and ensuring ethical considerations, equality, and promoting diversity, among other objectives. They are responsible for creating the policies and procedures to adopt this framework.

*2) Faculty and Committees:* They include the academic board, the Learning & Teaching committee, the Course/Program Coordinators and Unit Coordinators. Their aim is to stabilise the framework so that it provides fair evaluation, does not overburden the teachers, ensures the effectiveness of the curriculum, and provides enough learning opportunities for everyone without any bias and diversity concerns.

*3) Students:* Students need to be aware of the state-of-the-art GenAI tools and technologies. They should be able to understand the effective usage of these tools to utilise them to their fullest. Additionally, students need to be made aware of the ethical concerns and the power of GenAI.

*4) ICT Team:* The Information and Communication Technology (ICT) staff need to be upskilled and be aware of the changing environment and newer tools to effectively manage GenAI related changes in the academic curriculum and system requirements.

*5) External Parties:* Government and regulators, in collaboration with the technology and industry sectors, provide nationwide policies and guidelines that shape the future of GenAI in education. Together, they play a pivotal role in establishing ethical standards, promoting innovation, and ensuring equitable access to AI-powered tools and resources.

Next, we introduce a framework for the adoption of GenAI in education, enabling the stakeholder to articulate and communicate their intentions, followed by a process to facilitate curriculum design, teaching and evaluation mechanisms, etc.

### B. 4E Framework

We propose a **4E** (**E**mbrace, **E**nable, **E**xperiment, and **E**xploit) framework (Fig. 2) for these stakeholders to facilitate the incorporation of necessary updates during the adoption process. The proposed framework defines an iterative process to identify impacted areas, convey the clarity of changes required by/for everyone followed by monitor, measure and update based upon these observations. This framework is agnostic to the sizes of academic institutions/universities. Different phases in the framework provide guidelines about the process and tasks involved to successfully adopt GenAI into the scholarship of teaching and learning and help identify its impact on research.

*1) Embrace:* This phase of the adoption framework focuses on articulating the intent of adopting GenAI in the education system. Decision makers, including university top management, leadership, and academic committee representatives, must reach a consensus and formally document this intent within their policy framework [34]. This phase necessitates their recognition of GenAI's impact on curriculum redesign, faculty and student upskilling, and collaboration with external parties. As part of this process, decision-makers may find it necessary to redefine the vision and mission of the education framework to incorporate the use of GenAI and to craft new policies that empower faculty, coordinators, and moderators to experiment and enhance their methods accordingly.

This phase, in turn, results in an updated vision and policy for the university, along with a written statement of intent for faculty members and students regarding the adoption of GenAI into the university's academic framework. This clear objective will guide the Academic Board, Course Coordinators, and Unit Coordinators in incorporating GenAI as an integral component of their curriculum design and teaching methodologies. The *Embrace* phase marks the inception of a new process for the adoption of GenAI in education, followed by the *Enable* phase.

*2) Enable:* This phase constitutes the core of GenAI adoption within an academic ecosystem, where educators must identify the requirements and develop a plan for integrating GenAI into their curriculum. Course coordinators should analyse their curriculum for necessary changes in the whole course whereas unit coordinators must reevaluate and redefine learning outcomes, evaluation methodologies, and assessments to accommodate potential GenAI usage by both faculty and students. Academic staff, including heads of schools, course coordinators, unit coordinators, and faculty, require upskilling with the necessary tools and training for its use. With increased adoption, evaluation methods for assignments and exams must also become GenAI-friendly, necessitating significant exposure and learning of tools and technologies for both academic staff and students. Furthermore, this adoption also raises ethical, privacy, and academic integrity concerns [1]. Therefore, policies and guidelines regarding ethics, privacy, and academic integrity must be updated to address these issues. Faculty and students should be educated about ethical and privacy issues related to GenAI and always uphold academic integrity according to established guidelines.

The potential outcomes of this phase include documents outlining ethics, privacy, and academic integrity guidelines; a process to upskill faculty and students; an AI-friendly curriculum for learning and evaluation; and an ***Academic eValuation Matrix (AVM)*** for assessing the impact and effectiveness of these changes. *AVM*, which can incorporate both qualitative and quantitative measures, is not a fixed matrix and should be customised by the Academic Board and the Learning & Teaching Committee to align with the specific requirements of courses and units. The parameters for *AVM* ought to be established in line with their outcome objectives and the vision as outlined in the *Embrace* phase. These parameters might include indicators such as student and faculty awareness, satisfaction indices, required ICT efforts, readiness of students and faculty, campus placements, incidents or issues related to academic integrity, ethical usage, and privacy, etc. [29], [35]–[40]. An example *AVM* and how its corresponding parameters can be collected is presented in the *Appendix*. The *AVM* (Table A) includes the possible parameters, associated actors, and their description. These parameters can be collected, summarised, and prioritised through student and faculty surveys. A detailed approach to selecting these parameters, obtaining their values and further analysis is beyond the scope of this paper.

*3) Experiment:* This phase aims to use and experiment with the developed course and unit curriculum, along with the

tools and techniques learnt during previous phases. Unit coordinators play a crucial role in designing their respective units, modifying the evaluation approach, and providing feedback on the effectiveness and challenges faced during the adoption process [37]. The evaluation matrix (*AVM*) developed during the *Enable* phase can be used to measure and analyse the effectiveness of the modified teaching and learning methodologies, the effectiveness and usefulness of the tools and training, and identify any missing gaps. Based on the outcomes of these analyses and the learnings from the experiment, this phase provides feedback to the *Enable* phase. The completed evaluation matrix (*AVM*) is the outcome of this phase and is used by the academic team in the *Enable* phase to improve further based upon these learnings. The interaction between *Enable* and *Experiment* phases is an iterative process. This process continues until all stakeholders in both phases are satisfied with the outcome. Various courses and units encompassing different academic levels (entry level and advanced level), types of activities (tutorials, labs, lectures etc.) have different learning outcomes and may require different GenAI strategies. Therefore, during this phase, the course and unit coordinators may choose different course types, class sizes, units/subjects, and types of activities accordingly. This will ensure that these variabilities are accommodated and facilitate the emergence of a holistic view and best practices for diverse scenarios in the university setup. It should be noted that based on these diverse requirements, the *AVM* parameters can be adjusted accordingly. The lessons learnt and the methods employed during the iteration phase are leveraged to establish guidelines and standards. Subsequently, they can roll out the learnt processes and move to the *Exploit* phase for university-wide adoption.

*4) Exploit:* This marks the scaling-up phase of the framework. In this phase, faculty and students are exposed to the updated curriculum, along with the tools and techniques necessary to deliver high-quality education using GenAI. The ICT staff is involved in upgrading and expanding the systems to facilitate university-wide adoption. Course/Unit coordinators align their respective courses/units with insights from the *Enable* and *Experiment* phases. Throughout this process, the outcomes of each unit and course are monitored, and staff as well as students, can provide feedback and observations to the *Enable* Team. The evaluation matrix *AVM* can also be applied in this phase, and feedback can be provided to the teams involved in the *Enable* phase to provide further updates in the curriculum and training. However, this phase is expected to be a lengthier iteration in terms of time, typically requiring learning through at least three or more cycles of offering the same unit or course before providing feedback.

## VI. Conclusions and Future Directions

GenAI is a rapidly emerging field, requiring continuous updates to the curriculum and ongoing learning for both educators and students. To the best of our knowledge, this paper is the first comprehensive effort to identify the necessary stakeholders, their roles and responsibilities, and to provide standards and guidelines for the effective integration of GenAI into teaching, learning, and research. The proposed *4E* framework delineates various phases of adoption and offers a workflow to assist universities in adopting GenAI, ultimately enhancing the scholarship of teaching and learning, and also assessing its impact on research methodologies. To measure the impact of these changes, the framework also provides *AVM*, an evaluation matrix. These serve as a systematic approach to the integration of GenAI into the education ecosystem.

The next phase of this research involves validating and testing the framework in the university setting. The insights gained through experimentation and testing with the proposed framework can be used to improve information flow and university processes, ensuring the successful adoption of GenAI.

## Acknowledgments

We thank the Melbourne Institute of Technology (MIT) administration for providing financial and administrative support.

## Biography Section

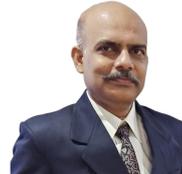

**Samar Shailendra** (Senior Member, IEEE) received his PhD from IIT Guwahati. He is currently working as a Senior Lecturer at the Melbourne Institute of Technology, Australia. He is also the Adjunct Professor at IIIT Bangalore. Previously, he worked with Intel and led their Mobile Edge Computing (MEC) Standardisation at 3GPP and India. He also served as the chair of the TSDSI Roadmap Committee and vice chair of Study Group - Networks (SGN) at TSDSI. He is credited with making substantial contributions to shaping 5G Advanced requirements and the global vision for 6G technology in close collaboration with Indian Telecom Standard Bodies and Government Agencies. His research interests include SDN/NFV, Internet Architecture, Transport Protocols, M2M, Drones, Robotics, AI and Quantum Computing.

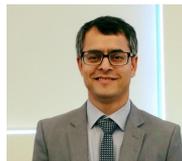

**Rajan Kadel** received the B.Eng. degree in computer engineering from the Tribhuvan University, Nepal, in 2002, the M.Sc. degree in telecommunications engineering from the University of Gävle, Sweden, in 2007, and the PhD Degree in telecommunications engineering from the University of South Australia (UniSA), Australia, in 2013. Currently, he holds the position of senior lecturer and course coordinator at the Melbourne Institute of Technology, Australia. He has more than a decade of teaching, research and professional experience. Previously, he worked for telecommunication sectors in different capacities. He worked as a switching supervisor for Nepal Telecom and assistant manager for Nepal Telecommunications Authority. His research interests cover learning and teaching, error-control coding, Wireless Sensor Network (WSN), and Wireless Body Area Network (WBAN).

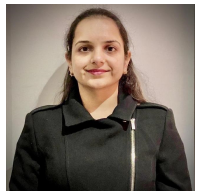

**Aakanksha Sharma** received her PhD from Federation University, Australia, in 2022. She currently serves as a Lecturer at Melbourne Institute of Technology. She has a strong teaching background and worked at Federation University and Royal Melbourne Institute of Technology (RMIT). Prior to this, she worked as an Assistant Professor at Chandigarh University for five years. Her research interests include the Internet of Things, Software-Defined Networks, wireless communications, Artificial Intelligence, and Quantum Computing.

## Appendix

Table A provides a sample *AVM* to be created and used to measure the effectiveness of GenAI adoption. This table contains the measurement parameters, the potential indicators of GenAI adoption, and the corresponding actors involved for each measurement criteria. These parameters encompass both qualitative and quantitative aspects. The subsequent columns provide descriptions of each parameter and possible approach(es) to measure them.

TABLE A
AN EXAMPLE OF ACADEMIC EVALUATION MATRIX (AVM)

| Parameters | Actors | Description | How to Measure |
|---|---|---|---|
| Awareness | Academic staff<br>Student<br>Administrative staff | This parameter measures the level of awareness among staff and students about GenAI technologies, their potential applications, and implications. | Measured through Surveys and Interviews. |
| Readiness | Academic staff<br>Student<br>Administrative staff | This parameter focuses on the preparedness of staff and students to adopt and effectively use GenAI technologies. | Evaluated through pre and post implementation surveys. |
| Ethics and Privacy | Academic staff<br>Student<br>Administrative staff | This parameter evaluates the ethical use of GenAI technologies, including adherence to privacy laws and respect for intellectual property. This also includes the management of privacy concerns associated with the use of GenAI, data protection and confidentiality. | Measured through the surveys and number of sessions conducted. |
| ICT Adoption and Automation | Academic staff<br>Student<br>Administrative staff | This parameter evaluates availability of GenAI tools and technologies to students, academic and administrative staff. This also covers the adoption of GenAI for system automation in the organisation. | Evaluated through pre and post implementation surveys. |
| Equitable Access | Academic staff<br>Student<br>Administrative staff | This parameter measures the accessibility of GenAI technology for everyone in the institute especially with respect to other similar organisations and geographies. | Measured through survey and interviews. |
| GenAI Placements | Students | This parameter evaluates the (change in) number of student placement for units incorporated with GenAI. | Measured through audit. |
| AI Placements | Students | This parameter evaluates the (change in) number of student placement in the area of AI. | Measured through audit. |
| Incidents of Academic Integrity | Academic staff<br>Students | This parameter evaluates the number of academic integrity incidents related with GenAI covering in both professional and academic setting. | Measured through audit. |
| Ethics and Privacy Incidents | Academic staff<br>Students<br>Administrative staff | This parameter evaluates the number of ethical and privacy related incidents related with GenAI in both professional and academic setting. | Measured through audit. |
| R & D Artefacts | Academic staff<br>Students | This parameter evaluates the number of research outputs (papers, patents, whitepapers, reports etc.). | Measured through audit. |
| R & D Collaborations | Academic staff<br>Students | This parameter evaluates the number of collaborations with industry, other local and international educational institutes and/or any other bodies. | Measured through audit. |
| Professional Development and Training | Academic staff<br>Students<br>Administrative staff | This parameter evaluates the number of professional development activities and training on GenAI tools and technologies to students, academic and administrative staff. | Measured through audit. |
| Participation | Academic staff<br>Students<br>Administrative staff | Student participation measures engagement, involvement in educational activities (such as class attendance, active participation, and assignment completion rate), commitment and interest in learning, providing insights into the effectiveness of the adopted teaching methods. | Measured through attendance record, survey and interviews. |
| Retention | Academic staff<br>Students<br>Administrative staff | This measures the percentage of student/staff continuing to the next academic year and/or level. | Measured through data analysis. |
| Progression | Students | The progression evaluates students' advancement through unit/course (can include factors such as completion rates, grades, and time to complete unit/course). | Measured through data analysis. |